\documentclass[twocolumn,prl]{revtex4}

\usepackage{graphicx}
\usepackage{mathptmx} 

\begin{document}

\noindent
{{\bf Comment on ``Earthquakes Descaled''}}

\vspace{0.3cm}

Lindman {\it et al.} \cite{Lindman} have used a nonhomogeneous Poisson 
process with a modified Omori rate, $r(t) \equiv dn/dt = r_M (1+t/c)^{-p}$,
to model earthquake 
occurrence.
We are going to show that 
contrary to claims in Ref. \cite{Lindman},
this extremely simple description is incomplete in order to explain
Bak {\it et al.}'s unified scaling law \cite{Bak}.

We generalize Lindman {\it et al.}'s model by introducing
an $r-$dependent waiting-time probability density of the form
$D(\tau |r) \propto r^\gamma \tau^{\gamma-1}e^{-r\tau /a}$,
which includes the nonhomogeneous Poisson process of Ref. \cite{Lindman},
given by $\gamma=1$ and $a=1$ 
(both parameters linked by normalization).
The probability density of the waiting times in the Omori sequence,
independent of $r$,
is given by the mixing of all $D(\tau|r)$ \cite{Corral03},
\begin{equation}
D(\tau | r_m) = \frac{1}{\mu}\int_{r_m}^{r_M} r D(\tau | r) \rho(r) dr 
\end{equation}
where 
$\rho(r)$ is the density of rates, $\rho(r) \propto |dr/dt|^{-1}$
$= C/r^{1+1/p}$;
$\mu$ is the mean rate of the sequence,
$\mu \equiv \int r \rho dr$;
$r_M$ is the maximum rate, corresponding to $t=0$; 
and
$r_m$ is the minimum rate,
related to the background seismicity level.
Note that we have emphasized the dependence on $r_m$.

Easy to deal with but illuminating is the case $\gamma=1/p$,
which yields 
\begin{equation}
D(\tau|r_m) \propto \frac {C}{ \mu} \, 
\frac{(e^{-r_m \tau /a}-e^{-r_M \tau/a})}{ \tau^{2-1/p}},
\end{equation}
where the minimum rate $r_m$ determines the exponential tail
of $D(\tau|r_m)$ for large $\tau$,
preceded by a decreasing power law with exponent $2-1/p$ if $r_M \gg r_m$.
For $p=1$ this
is in agreement with the simulation and numerics 
in Fig. 1 of Ref. \cite{Lindman};
however, it can be shown that the exponent $2-1/p$
holds even when $\gamma \ne 1/p$, which is in disagreement
with Lindman {\it et al.}'s claim of a $1/\tau^p$ decay for $p<1$
and $1/\tau^{\sqrt{p}}$ for $p>1$.

Nevertheless, this description totally ignores the spatial 
degrees of freedom, fundamental in Bak {\it et al.}'s approach.
In fact, their approach performs a mixing 
of waiting times coming from different
spatial areas (or cells), which are characterized by 
disparate seismic rates.
In particular, each area will have a different
$r_m$, depending on its background seismicity level.
This spatial heterogeneity of seismicity can be described by a 
power-law distribution of mean rates $R$,
being $R$ the total number of events divided by the total time for 
a given area,
see Fig. 1 and Ref. \cite{Corral03};
if we assume that the minimum rate $r_m$ 
is directly related to the mean rate of the sequence $\mu$,
which in turn is in correspondence with the mean rate
in the area, $R$,  
then,
$p(r_m)\propto 1/r_m^{1-\alpha}$
and therefore the waiting-time probability density comes from
the mixing,
\begin{equation}
D(\tau) \propto \int_{r_{mm}}^{r_{mM}} r_m D(\tau|r_m) p(r_m)dr_m
\end{equation}
where $r_m$ varies between $r_{mm}$ and $r_{mM}$.
Integration, taking into account that $C/\mu$ depends on $r_m$,
leads to
\begin{equation}
D(\tau) \propto
1 / {\tau^{2+\alpha}} 
\quad \text{ for } \quad r_{mm}\tau \ll 1 \ll r_{mM}\tau,
\label{2ndpower}
\end{equation}
which is in disagreement with Lindman {\it et al.}'s analysis.

In fact, the power law for long times [Eq. (\ref{2ndpower})], was established 
in Ref. \cite{Corral03} for Southern California,
but without relating it to the spatial heterogeneity of seismicity.
The universal value of the exponent $2+\alpha$, 
found in Ref. \cite{Corral_physA} analyzing diverse seismic catalogs,
would imply the universality of seismicity spatial heterogeneities.
In consequence, Bak {\it et al.}'s unified scaling law provides a
way to measure these properties and is far from being as trivial as
suggested by Lindman {\it et al.}'s approach.

The model presented here is still too simple for real seismicity,
but provides a clear visualization of its complexity 
and the fundamentals of the unified scaling law of earthquakes.


\vspace{0.3cm}

\noindent
\'Alvaro Corral$^*$ 
and Kim Christensen$^\dag$

\noindent
$^*$%
Departament de F\'\i sica, 
Universitat Aut\`onoma de Barcelona,
E-08193 Bellaterra, Spain;
$^\dag$%
Blackett Laboratory, Imperial College, 
Prince Consort Road, London SW7 2BW, UK.

\today

PACS numbers: 91.30.Dk, 05.65.+b, 89.75.Da



\begin{figure}
\centering
\includegraphics[width=3.5in]{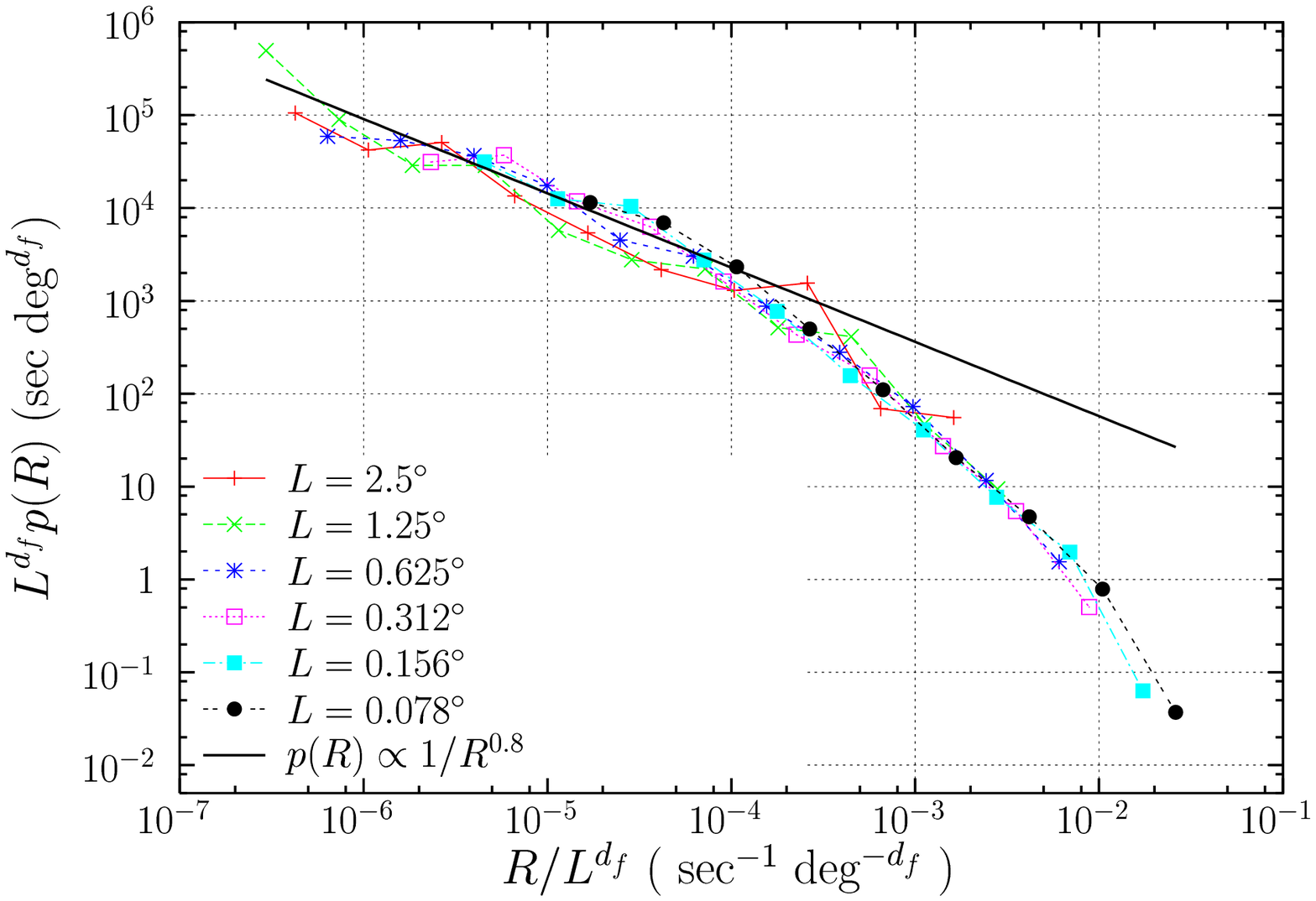}
\caption{
(color online)
Distribution of mean rates $R$ for earthquakes with magnitude $M \ge 2$
in Southern-California,
dividing the area $(123^\circ W, 113^\circ W) \times (30^\circ N, 40^\circ N)$ 
in cells of size $L$, ranging from $L=0.078^\circ$ to $L=2.5^\circ$, and
averaging the periods 1984-1992 and 1993-2001.
The distributions are rescaled by $L^{d_f}$ with $d_f=1.6$. 
For small $R/L^{d_f}$, the data are in agreement with a density
$p(R) \propto 1 / R ^{0.8}$, obtained from a fit.
\label{Dr}
}
\end{figure}


\begin{thebibliography}{99} 

\bibitem{Lindman}
M. Lindman {\it et al.},
{Phys. Rev. Lett.} {\bf 94}, 108501 (2005).

\bibitem{Bak}
P. Bak {\it et al.},
{Phys. Rev. Lett.} {\bf 88}, 178501 (2002).

\bibitem{Corral03}
A. Corral,
Phys. Rev. E, {\bf 68}, 035102 (2003).

\bibitem{Corral_physA}
A. Corral,
Physica A {\bf 340}, 590 
(2004). 







\end{thebibliography}
\end{document}